\documentclass{sigchi}

\usepackage{balance}       
\usepackage{graphics}      
\usepackage[T1]{fontenc}   
\usepackage[utf8]{inputenc}
\usepackage{txfonts}
\usepackage{mathptmx}
\usepackage[pdflang={en-US},pdftex]{hyperref}
\usepackage{color}
\usepackage[table]{xcolor}
\usepackage{booktabs}
\usepackage{csquotes}
\usepackage{textcomp}
\usepackage{enumitem}
\usepackage{ntheorem}
\usepackage{amsmath}
\usepackage[LGRgreek]{mathastext}

\usepackage{microtype}        
\usepackage{ccicons}          
\usepackage{todonotes}
\usepackage{comment}

\def\plaintitle{SIGCHI Conference Proceedings Format}

\def\emptyauthor{}
\def\plainkeywords{Authors' choice; of terms; separated; by
  semicolons; include commas, within terms only; required.}

\makeatletter
\def\url@leostyle{%
  \@ifundefined{selectfont}{
    \def\UrlFont{\sf}
  }{
    \def\UrlFont{\small\bf\ttfamily}
  }}
\makeatother
\def\pprw{8.5in}
\def\pprh{11in}

\definecolor{linkColor}{RGB}{6,125,233}
\hypersetup{%
  pdftitle={\plaintitle},
  pdfauthor={\emptyauthor},
  pdfkeywords={\plainkeywords},
  pdfdisplaydoctitle=true, 
  bookmarksnumbered,
  pdfstartview={FitH},
  colorlinks,
  citecolor=black,
  filecolor=black,
  linkcolor=black,
  urlcolor=linkColor,
  breaklinks=true,
  hypertexnames=false
}
\begin{document}

\title{Crowd Guilds: Worker-led Reputation and Feedback\\ on Crowdsourcing Platforms}

 \numberofauthors{1}
 \author{ 
   \alignauthor{Mark E. Whiting, Dilrukshi Gamage, Snehalkumar (Neil) S. Gaikwad, Aaron Gilbee, Shirish~Goyal, Alipta Ballav, Dinesh Majeti, Nalin Chhibber, Angela Richmond-Fuller, Freddie~Vargus, Tejas Seshadri Sarma, Varshine Chandrakanthan, Teogenes~Moura, Mohamed~Hashim~Salih, Gabriel~Bayomi~Tinoco~Kalejaiye, Adam Ginzberg, Catherine~A.~Mullings, Yoni Dayan, Kristy~Milland, Henrique Orefice, \\Jeff Regino, Sayna Parsi, Kunz Mainali, Vibhor Sehgal, Sekandar Matin, \\Akshansh Sinha, Rajan Vaish, Michael S. Bernstein\\
    \affaddr{Stanford Crowd Research Collective}\\
     \email{daemo@cs.stanford.edu}}\\
}

\maketitle
\begin{abstract}
Crowd workers are distributed and decentralized. While decentralization is designed to utilize independent judgment to promote high-quality results, it paradoxically undercuts behaviors and institutions that are critical to high-quality work. Reputation is one central example: crowdsourcing systems depend on reputation scores from decentralized workers and requesters, but these scores are notoriously inflated and uninformative. In this paper, we draw inspiration from historical worker guilds (e.g., in the silk trade) to design and implement \textit{crowd guilds}: centralized groups of crowd workers who collectively certify each other's quality through double-blind peer assessment. A two-week field experiment compared crowd guilds to a traditional decentralized crowd work model. Crowd guilds produced reputation signals more strongly correlated with ground-truth worker quality than signals available on current crowd working platforms, and more accurate than in the traditional model.
\end{abstract}

\keywords{crowdsourcing platforms; human computation}
\category{H.5.3.}{Group and Organization Interfaces}{}

\section{Introduction}
Crowdsourcing platforms such as Amazon Mechanical Turk decentralize their workforce, designing for distributed, independent work~\cite{gray2016crowd, mcinnis2016taking}. Decentralization aims to encourage accuracy through independent judgement~\cite{surowiecki2005wisdom}. However, by making communication and coordination more difficult, decentralization disempowers workers and forces worker collectives off-platform~\cite{martin2014being, yin2016communication, gray2016crowd}. The result is disenfranchisement~\cite{irani2013turkopticon, salehi2015we} and an unfavorable workplace environment~\cite{martin2014being, mcinnis2016taking}. Worse, while decentralization is motivated by a desire for high-quality work, it paradoxically undercuts behaviors and institutions that are critical to high-quality work. In many traditional organizations, for example, centralized worker coordination is a keystone to behaviors that improve work quality, including skill development~\cite{billett2001learning}, knowledge management~\cite{lee2003knowledge}, and performance ratings~\cite{sparrowe2001social}.

\begin{figure}[t!]
 \centering
 \includegraphics[width=\columnwidth]{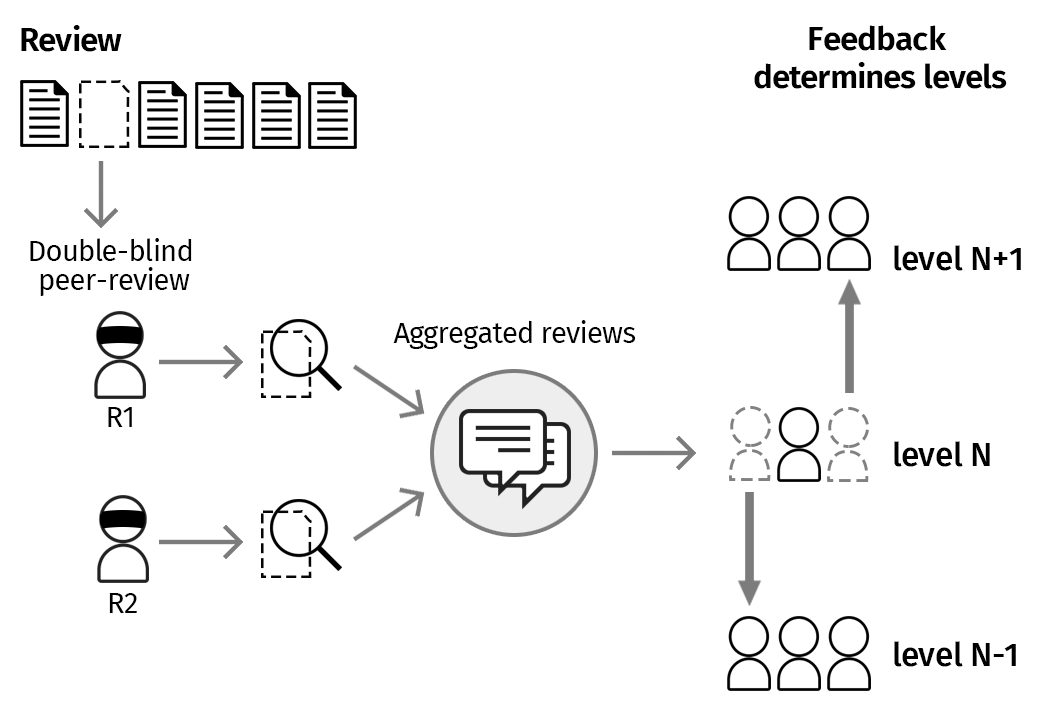}
 \caption{Crowd guilds provide reputation signals through double blind peer-review. The reviews determine workers' levels.}
 \label{fig:crowdguildconcept}
\end{figure}

In this paper, we focus on \textit{reputation} as an exemplar challenge that arises from worker decentralization: effective reputation signals are traditionally reliant on centralized mechanisms such as performance reviews~\cite{sparrowe2001social, josang2007survey}. Crowdsourcing platforms rely heavily on their reputation systems, such as task acceptance rates, to help requesters identify high-quality workers ~\cite{irani2013turkopticon, mitra2015comparing}. On Mechanical Turk, as on other on-demand platforms such as Upwork and Uber, these reputation scores are derived from decentralized feedback from independent requesters. However, the resulting reputation scores are notoriously inflated and noisy, making it difficult for requesters to find high-quality workers and difficult for workers to be compensated for their quality~\cite{mitra2015comparing, horton2015reputation}.

To address this reputation challenge, and with an eye toward other challenges that arise from decentralization, we draw inspiration from a historical labor strategy for coordinating a decentralized workforce: \textit{guilds}. Worker guilds arose in the early Middle Ages, when workers in a trade such as silk were distributed across a large region, as bounded sets of laborers who shared an affiliation. These guilds played many roles, including training apprentices~\cite{guthrie2007learning, mocarelli2008guilds}, setting prices~\cite{ogilvie2004guilds}, and providing mechanisms for collective action~\cite{renard1918guilds, perez2008inventing}. Especially relevant to the current challenge, guilds measured and certified their own members' quality~\cite{guthrie2007learning}. While guilds eventually lost influence due to exerting overly tight controls on trade~\cite{ogilvie2004guilds} and exogenous technical innovations in production, their intellectual successors persist today as professional organizations such as in engineering, acting and medicine~\cite{ogilvie2005use, larson1979rise}. Malone first promoted a vision of online ``e-lancer'' guilds twenty years ago~\cite{malone1999will}, but to date no concrete instantiations exist for a modern, online crowd work economy.

We present \textit{crowd guilds}: crowd worker collectives that coordinate to certify their own members and perform internal feedback to train members (Figure~\ref{fig:crowdguildconcept}). Our infrastructure for crowd guilds enables workers to engage in continuous double-blind peer assessment~\cite{kulkarni2015peer} of a random sample of members' task submissions on the crowdsourcing platform, rating the quality of the submission and providing critiques for further improvement. These peer assessments are used to derive guild levels (e.g., Level 1, Level~2) to serve as reputation (qualification) signals on the crowdsourcing platform. As workers gather positive assessments from more senior guild members, they rise in levels within the guild. Guilds translate these levels into higher wages by recommending pay rates for each level when tasks are posted to the platform. While crowd guilds focus here on worker reputation, our experiment implementation also explores how crowd guilds could address other challenges such as collective action (e.g., collectively rejecting tasks that pay too little), formal mentorship (e.g., repeated feedback and training), and social support (e.g., on the forums). Because existing platforms cannot be used to support these affordances, we have implemented crowd guilds on the open-source Daemo crowdsourcing platform~\cite{gaikwad2015daemo}.

As with historical guilds, several crowd guilds can operate in parallel across different areas of expertise. To focus our exploration, our prototype in this paper implements a single platform-wide guild for microtask workers. While in theory anybody can perform microtask work, prior work has demonstrated that it requires a wide range of both visible and invisible expertise to perform supposedly `simple' work effectively~\cite{martin2014being}. This need for expert, high-quality microtask work has previously driven both platforms and requesters to curate private groups of expert workers (e.g., Mechanical Turk Masters, requesters' private qualification groups)~\cite{mitra2015comparing}. In our case, this expertise makes microtask work an appropriate candidate for a guild.

We performed a two-week field experiment to evaluate whether crowd guilds establish accurate worker reputation signals. We recruited 300 workers from Amazon Mechanical Turk, randomizing them between control and crowd guild conditions. We then launched tasks daily to the workers in these two groups, providing both conditions with a forum and automatically-generated peer assessment tasks. In the crowd guild condition, the peer assessment tasks determined guild levels and workers received the assessment feedback. In the control condition, no guild levels were available, and the peer assessments were never returned to the worker being assessed. We calculated each worker's accuracy using gold-standard tasks launched on the platform. 

Guilds' peer assessed ratings were a significantly better predictor of workers' actual accuracy than the workers' acceptance rates on Amazon Mechanical Turk. Furthermore, workers' peer assessment was significantly more accurate and less inflated in the guilds condition, when peers' reputations were attached to the assessment. Workers in the guilds condition also provided one another with more actionable feedback and advice than those in the control condition.

In sum, this paper contributes a design and infrastructure for crowd guilds as a re-centralizing force for crowdsourcing marketplaces. We target crowd guilds at addressing reputation challenges, because reliable reputation is a difficult and representative problem of the negative outcomes of worker decentralization. To follow, we review related work on crowd collectives and historical guilds, then introduce our design and describe our field experiment.

\section{Related work}
In this section, we first review literature on how crowdsourcing marketplaces can improve work quality, focusing on peer assessment methods. We then draw upon literature on the historical formation of guilds and their legacy to the modern world with a focus on structures for crowd workers to enhance their work reputation and improve communication in the online crowd work environment. Finally we discuss literature on crowdsourced worker communities and their collective behavior in online labor markets.

\subsection{Improving work quality} 
Ensuring high-quality crowd work is crucial for the sustainability of a microtask platform. Mechanisms such as voting by peer workers~\cite{callison2009fast}, establishing agreement between workers~\cite{sheng2008get}, and machine learning models trained on low-level behaviors~\cite{rzeszotarski2011instrumenting}, have been used to gauge and enhance the quality of crowd work. In addition to these techniques, task-specific feedback help crowd workers augment their behavior and improve their performance~\cite{dow2012shepherding}.

Crowdsourcing platforms have collected work feedback through requesters, workers, using self-assessment rubrics, and with the help of expert evaluators. The Shepherd system~\cite{dow2012shepherding} allows workers to exchange synchronous feedback with requesters. Crowd guilds scale up this notion of distributed feedback~\cite{campbell2015thousands} to make peer assessment and collective reputation management a core feature of a crowdsourcing platform.

Self-assessment is another route to help workers reflect, learn skills and more clearly draw connections between learning goals and evaluation criteria~\cite{boud2013enhancing}. However, workers in self-assessment systems become dependant on rubrics or use special domain language, which tends to be difficult for novices to understand~\cite{boud2000sustainable}. Automated feedback~\cite{haas2015argonaut} also enhances workers' performance. However, such systems are generally used to enhance the capabilities of specialized platforms; for example, Duolingo and LevelUp integrate interactive tutorials to enhance the quality of work~\cite{dontcheva2014combining}, which requires significant customization for a given domain, and has not been demonstrated in general work platforms.

\subsection{Peer assessment driving quality}
If crowd workers can effectively assess each other, they could bootstrap their own work reputation signals~\cite{zhu2014reviewing, kulkarni2015peer}. Worker peer review can scale more readily than external assessment, and leverages a community of practice built up amongst workers with the same expertise~\cite{dow2012shepherding}. It can also be comparably accurate: peer assessment matches expert assessment in massive open online classes~\cite{kulkarni2015peer}. 

For effectively assessing each other's contributions, it may be prudent to recruit assessors based on prior task performance. Algorithms can facilitate this by adaptively selecting particular raters based on estimated quality, focusing high quality work where it's most needed~\cite{peng2010decision}. The Argonaut system~\cite{haas2015argonaut} and MobileWorks~\cite{kulkarni2012mobileworks} demonstrate reviews through hierarchies of trusted workers. However, promotions between hierarchies in the Argonaut system require human managers and MobileWorks needs managers to play an essential procedural role in the quality assurance process by determining the final answer for tasks that are reported by workers as difficult, which restricts these systems' ability to scale. In contrast, crowd guilds provide automatic reputation levels based on peer assessment results to algorithmically establish who is qualified to evaluate which work.

The advice and feedback provided in peer assessment can facilitate distributed mentorship~\cite{campbell2015thousands}. Self and peer assessment can train reviewers to become better at the craft themselves~\cite{kulkarni2015peer, anderson2011crowdsourcing, dow2012shepherding}, and feedback from those with more expertise improves result quality~\cite{dow2012shepherding, luther2015structuring}. Returning peer assessment feedback rapidly can increase iteration and aid learning toward mastery~\cite{kulkarni2015peerstudio}. In crowd guilds, we introduce a review framework that draws on these insights, using workers' assessments of their peers' work to offer quality based reputation categories and constructive feedback.

Building on the above approaches, crowd guilds utilize the observation that crowd members can evaluate each others' work accuracy~\cite{dow2012shepherding, haas2015argonaut}. Additionally, we demonstrate that crowd guilds can create not just a rating for individual pieces of work, but a stable and informational reputation system.

\subsection{Guilds} 
Existing crowdsourcing markets deploy ad hoc methods for improving and sustaining the community of workers. To design a holistic community, we take inspiration from \textit{guilds}. Historically, guilds represented groups of workers with shared interests and goals, as well as enabling large-scale collective behaviors such as reputation management.

Guilds originally evolved as associations of artisans or merchants who controlled the practice of their craft in medieval towns~\cite{perez2008inventing}. These craftspeople's guilds, formed by experienced and confirmed experts in their respective fields of handicraft, behaved as professional associations, trade unions, cartels, and even secret societies~\cite{rosser1997crafts}. They used internal quality evaluation processes to progress members through a system of titles, often starting with \textit{apprentices}, who would go through some schooling with the guild, to then become \textit{journeymen}~\cite{guthrie2007learning}, and eventually develop to \textit{master craftsmen}~\cite{mocarelli2008guilds}. Guilds prided themselves on their collective reputation, which was an aggregate of their members' progress through the quality system, and for high quality work, which enabled them to demand premium prices. Some guilds even fined members who deviated from the guild's quality standards~\cite{ogilvie2004guilds}.

Today, the intellectual inheritance of guilds persists via professional organizations, which replicate some of their benefits~\cite{kieser1989organizational}. Professions such as architecture, engineering, geology, and land surveying require varying lengths of apprenticeships before one can gain professional certification, which holds great legal and reputational weight~\cite{ogilvie2005use, larson1979rise}. However, these professions fall into traditional work models and do not cater for global and location-independent flexible employment~\cite{laubacher1997flexible}. Non-traditional work arrangements such as freelancing do not provide common work benefits such as economic security, career development, training, mentoring and social interaction with peers, which are legally essential for classification as full-time work in many cases~\cite{karoly200421st}. Although the support of professional organizations exists for freelance work, much of the underlying value of guilds does not.

Guilds re-emerged digitally in Massively Multiplayer Online games (MMOs) and behave somewhat like their brick-and-mortar equivalents~\cite{poor2015mmo}. Guilds help their players level their characters, coordinate strategies, and develop a collectively recognized reputation~\cite{ducheneaut2007life, kang2009impact}. This paper draws on the strengths of guilds in assessing participants' reputation, and adapts them to the non-traditional employment model of crowd work. Crowd guilds formalize the feedback and advancement system so that it can operate at scale with a distributed membership. 

Guilds offer an attractive model for crowd work because they provide reputation information for distributed professionals and carry a collective professional identity. They could also be used to train their own members~\cite{suzuki2016atelier, rees2015building, kaye2012putting}, and may eventually give workers the opportunity to take collective action in response to issues with requesters or the platform, all of which are characteristics notably missing from today's crowdsourcing ecosystem.

\subsection{Collective action} 
Despite being spread across the globe, crowd workers collaborate, share information, and engage in collective action~\cite{gray2016crowd, yin2016communication, gupta2014turk}. The Turkopticon activist system, for instance, provides workers with a collective voice to publically rate requesters~\cite{irani2013turkopticon}, though it remains a worker tool, external to the Mechanical Turk platform. Along with Turkopticon, workers frequent various forums  to share information identifying high-quality work and reliable requesters~\cite{martin2014being}. However, these forums are hosted outside the marketplace. This de-centralization from the main platform makes it harder to locate and obtain reputational information~\cite{josang2007survey} and, when needed, bring about large scale collective action. Therefore, Dynamo identified strategies for small groups of workers to incite change~\cite{salehi2015we}.

In existing marketplaces, workers are frustrated by capricious requester behavior related to work acceptance and having limited say in cultivating platform policies~\cite{deng2013crowdsourcing, kittur2013future}. Issues related to unfair rejections, unclear tasks, and platform policies have been publicly discussed~\cite{martin2014being}, but workers have limited opportunities to impact platform operations leaving less room to accommodate emerging needs. Therefore, crowd guilds focus on providing peer assessed reputation signals. To do this, they internalize affordances of previous tools for peer-review and gathering feedback, and give the guild power in the marketplace to manage some of these issues.

\section{Crowd guilds}
In this section, we describe technical infrastructure for \textit{crowd guilds} in a paid crowdsourcing marketplace, enabling collective evaluation of members' reputations via peer feedback.

A crowd guild is a group of crowd workers who coordinate to manage their own reputation. As a research prototype, we have implemented the guild structure with the Daemo open-source crowdsourcing platform~\cite{gaikwad2015daemo}. Crowd guilds could be built to focus on specific types of tasks and cultivate expertise for that labor. With a single platform-wide guild, however, we focus on the visible and invisible aspects of microwork, such as the expertise necessary to perform Mechanical Turk tasks effectively~\cite{martin2014being}. We thus designed the crowd guilds in active collaboration with workers on Amazon Mechanical Turk forums. Our discussions with workers on forums and over video chat led to several design decisions in crowd guilds (e.g., payment for peer assessment tasks).

Crowd guilds introduce a peer assessment infrastructure allowing workers to review each other's work, provide feedback via work critiques, and establish publicly-visible guild worker levels (e.g., Level 2, Level 3). To complement crowd guilds and explore the design space, we have also created prototype implementations of other behaviors that guilds can support: collective social spaces for informal engagement~\cite{yu2014comparison}, group determination of appropriate wage levels, and collective rejection of inappropriate work.

\subsection{Reputation via peer assessment and guild leveling}
Because each worker on a crowdsourcing platform is essentially a freelancer, reputation information (e.g., acceptance rates and five-star ratings) is generally uncoordinated and often highly inflated~\cite{kittur2013future, horton2015reputation}. Historically, in similar situations, guilds stepped in to guarantee the quality of their own (similarly independent) members, just as professional societies do today. Thus, crowd guilds form around a community of practice~\cite{wenger2011communities} that can effectively assess its own members' skills. Crowd guilds manage their members' reputation by assigning public guild levels to each member. To do so, our infrastructure randomly samples guild members' work, anonymizes it, and routes it to more senior members for evaluation to determine each worker's appropriate level.

\begin{figure}[tb]
 \centering
 \includegraphics[width=\columnwidth]{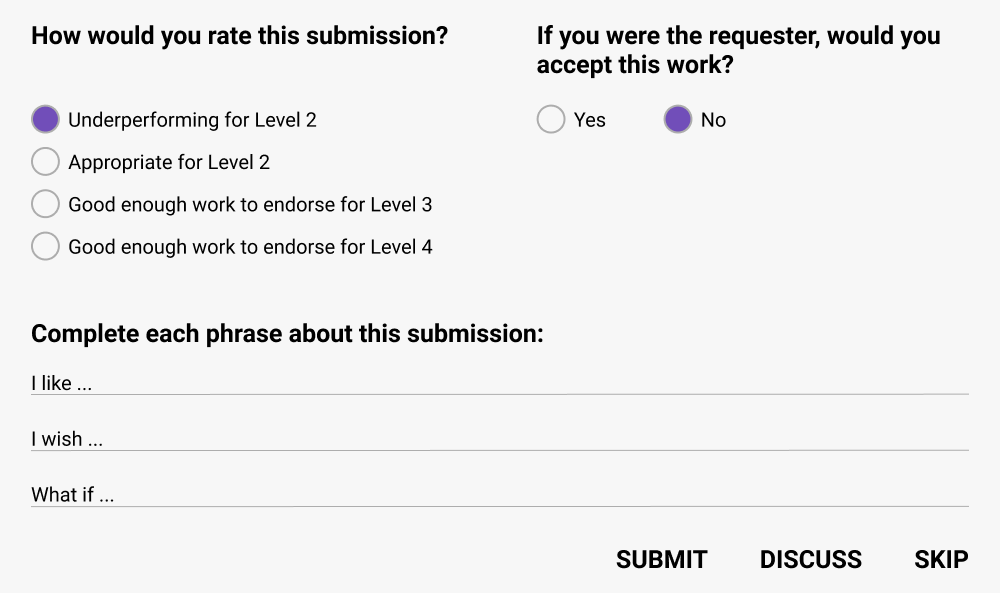}
 \caption{Crowd guilds peer assess a random sample of their own members' work in order to determine promotion between levels. The assessment form asks the double-blind reviewer to rate the work relative to the worker's current level, and give open-ended feedback.}
 \label{fig:review}
\end{figure}

\subsubsection{Peer review}
Our system automatically triggers double-blind peer reviews of tasks completed by guild workers on the crowdsourcing platform. Online peer assessment~\cite{kulkarni2015peer} is in regular use for filtering out low-quality work in crowdsourcing~\cite{little2010turkit, little2010exploring}---crowd guilds adapt this process for use in promotion and reputation. Peers and professionals are accurate at assessing each other~\cite{falchikov2000student, topping1998peer}, but people trust information more when it comes from those with more experience and authority~\cite{patel2012power}. Thus, a worker's peer reviews in a crowd guild can only be completed by guild members ranked one level above: for example, Level 2 workers can only be assessed by Level 3 workers.

To generate reviews, the crowd guilds infrastructure randomly samples a percentage of each worker's task submissions. Each sampled submission is wrapped inside an evaluation task and posted back onto the platform as a paid assessment task available to qualified members of the guild.

The reviews are double-blind---the workers do not know who reviewed their work, or which of their tasks will be reviewed, and the reviewers do not know which worker they are reviewing. As a result, workers cannot strategically increase their work effort to get more positive reviews, and reviewers have little opportunity to be selective about their reviews to benefit particular parties. Because reviews are conducted by guild members one level above the worker being reviewed, the quality expectations of the reviewer are not too distant from those of the worker, and when a guild member's level increases, the quality standards to proceed to the next level are proportionately higher. Quality standards are not explicitly set for any level, but are interpreted by those in the level as their reviews decide who will join them. 

Review tasks consist of a copy of the original task, the worker's response and three review questions (Figure~\ref{fig:review}):

\begin{figure*}[t!]
 \centering
 \includegraphics[width=.65\textwidth]{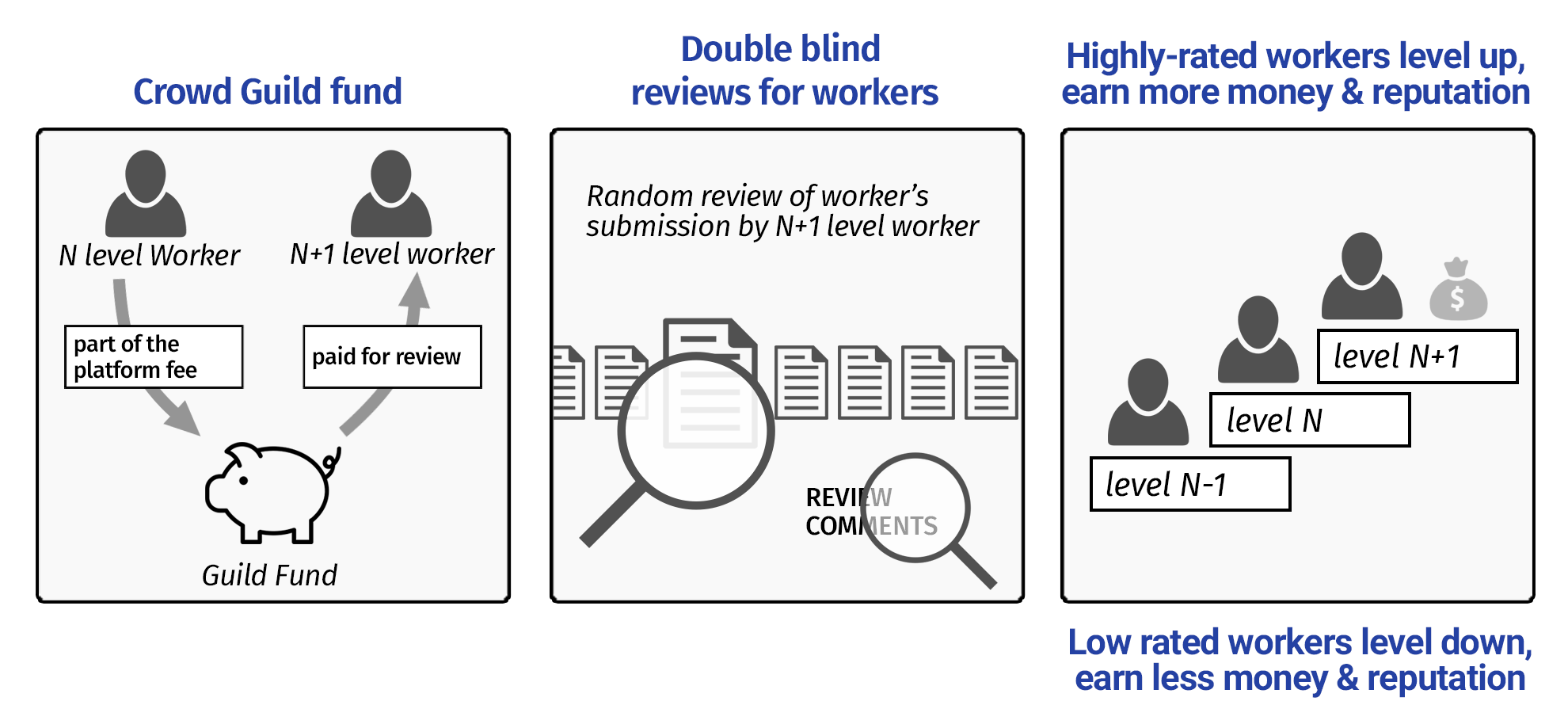}
 \caption{Review and leveling combine to form reputation ranks. Reviews are funded through the platform and are conducted, double-blind, on random tasks. The level of the worker is adjusted as the moving average of recent reviews meets thresholds. The moving average of recent reviews is used to adjust the level of the worker if they meet a threshold.}
 \label{fig:leveling}
\end{figure*}

\begin{enumerate}
\item A four-point ordinal scale for evaluating the quality of the work. The scale points are calibrated to the worker's current level~$n$, and ask the reviewer to rate the results as: a) Subpar: appropriate for level~$n-1$, b) At par: appropriate for level~$n$, c) Above par: appropriate for level~$n+1$, or d) Far above par: appropriate for level~$n+2$. Workers collectively develop criteria for each level as the guild matures. Over time, such ratings will determine whether the worker stays, moves up or moves down in the guild levels. The inclusion of a~$n+2$ assessment option was based on crowd worker feedback that it should be possible to rapidly level up after a skilled worker joins a guild. 
\item A question asking if the worker thinks the requester is likely to accept the work. Framing focusing on the requester's perspective (``If you were the requester...'') results in lower variance responses compared with asking for egocentric reviews (``Would you accept this?'')~\cite{gilbert2014if}. This question is not expected to be used to adjust actual acceptance of work, but gives useful insight into workers' perceptions of requesters' interests.
\item An open-ended text field asking how the worker might improve their work in the future. This field is designed to enable critique, utilizing a ``I Like, I Wish, What If'' model drawn from design thinking (e.g.,~\cite{ulibarri2014research}) because of its efficacy in motivating high quality responses that are actionable and prosocial.
\end{enumerate}

To make it less likely that senior guild members exercise unfair power, review tasks are also included in the tasks sampled for review. Reviews themselves get reviewed---a form of meta-reviewing. These meta-reviews are completed by all guild members, not just higher-ranked members. If a guild member is reviewing unfairly, others in the guild can recognize and punish the behavior. Meta-reviews also ensure that the quality standards for a level are reasonable.

Review tasks are paid tasks on the platform. Funds for review could come from the requesters, workers or platform, and could operate like a subscription, tax, or donation. Requesters are familiar with paying platform fees such as on Amazon Mechanical Turk, while platforms already invest in attracting workers and requesters; each stand to benefit from crowd guilds. However, the workers benefit most directly through increased wages as a result from leveling. Donation models can lack stability, and subscriptions can suffer from a lack of granularity, while a tax on individual tasks avoids these problems. In our design, we chose for crowd guilds to exact a marginal cost per task from worker earnings. In practice, this means that as workers complete tasks, funds will accumulate to pay for a review of one of those tasks, selected at random. 

Based on initial pilot analysis to establish the average time per review for several different task types, charging 10\% per task and conducting a review after 10 tasks have been completed is a default that serves both reviewers and workers well. Reviews for many task types are significantly faster than the original task. Because review tasks are fast, a 10\% overhead is enough for the more expensive higher-leveled workers to perform the review. However, an algorithm could be designed to tune these defaults dynamically, if task duration and review duration were measured by the platform. For example, in the long term, 10\% may in fact be oversampling the reviews, as workers perform hundreds of tasks per day. 

In practice, this system must be adjusted to deal with cold-start problems and privacy issues. If there are no workers of higher reputation (e.g., if a guild has just started or if a top-level worker is being reviewed), peers at the same level can review the task. In addition, some requesters do not want other workers seeing their tasks in order to protect private information. For this reason, Daemo requesters can opt out of their tasks being used for review purposes, which will avoid tasks being seen by anyone other than the initial worker, but this will also mean that additional information gained by the review process will not be returned to them. 

\subsubsection{Levels}
Levels (Figure~\ref{fig:leveling}) provide trusted reputation signals to requesters on the platform. All workers begin at Level 1.

\begin{figure}[tb]
 \centering
 \includegraphics[width=\columnwidth]{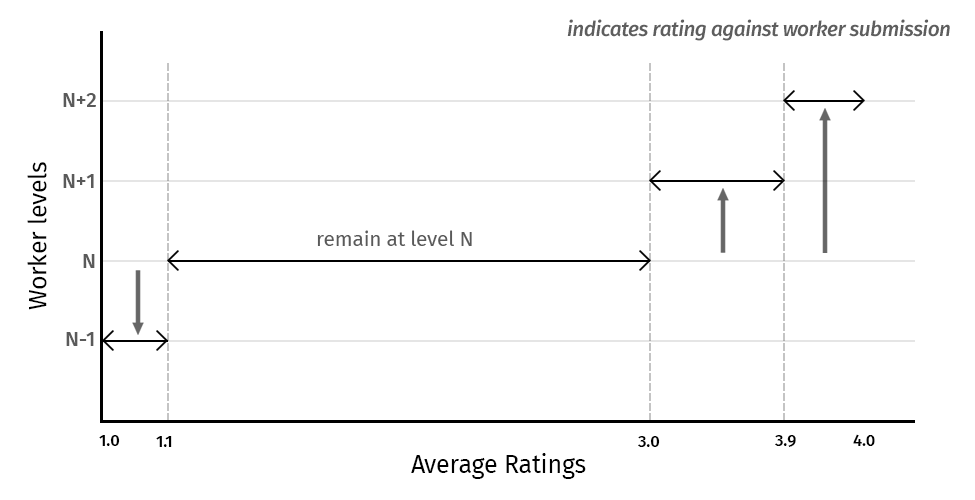}
 \caption{Example of leveling thresholds. The average review on a 1-4 scale, is used to choose one of 4 level shifts relative to the worker's current level: move down one level, remain in the same level, move up one level, or move up two levels.}
 \label{fig:thresholds}
\end{figure}

Promotion is determined using the peer review feedback. The four-point ordinal scale from peer review is converted into a numeric scale 1--4, and a moving numeric average is calculated across a window of 10 reviews. When a worker's moving average reaches a threshold such as those shown in Figure~\ref{fig:thresholds}, their level will be updated. After every level change, the moving average is reset. Unlike Mechanical Turk's acceptance rate, in which a blunder can mean a permanent mark on a worker's reputation, crowd guilds consider a moving average so as not to hold workers back by their past mistakes.
On the other hand, a worker can also be leveled down if they consistently produce low quality work. In this way, workers have an incentive to continue doing good work. 

The thresholds in Figure~\ref{fig:thresholds} only level down if work quality is particularly low, rated 1 out of 4 for 90\% of reviews. This set of thresholds has been used for our current deployment and has worked effectively as we are testing with a small number of workers, but the thresholds should be tuned to deal with large numbers of workers and large numbers of tasks to perform at scale.

Other approaches to leveling we considered include a periodic review schedule or a manual review board. A periodic review cycle ties workers to a timeline that is not related to actual changes in their quality and can move too slowly for active and skilled workers. A manual board voting on promotions, as in Wikipedia~\cite{leskovec2010signed}, would incur a larger cost overhead and increase the risk of an oligarchic regime managing the guild. Thus, we chose a continuous random sampling of work.

\subsection{Exploring social effects}
Historically, guilds did not restrict their attention to reputation management; they also engaged in a broader set of collective behaviors to support their trade and their members' welfare. While our attention in this paper is focused on reputation, crowd guilds can also explore other collective behaviors. We present several such prototypes: informal social engagement through collective spaces, group determination of wage levels, and collective rejection of inappropriate work.

\subsubsection{Level-based pricing}
If requesters can trust workers to be higher quality, then they can rely less on redundancy and pay individual workers more. When a requester posts a new task on Daemo and selects a guild level to target, higher levels show higher wage recommendations, based on the average hourly wage of workers at that level. The average hourly wage is calculated in aggregate over all members of a level, only counting work they perform that was posted to that level. This offers a useful insight into how much workers expect to make from a task---wage is often hard for requesters to make a judgement about without a trial and error process~\cite{cheng2015measuring}. Leaving this as a recommendation, and not a requirement, allows for some flexibility and for the pricing scheme to organically evolve. 

\subsubsection{Collective rejection}
Because some requesters routinely disregard ethical wage standards, crowd guilds also provide workers with a task rejection mechanism. Guild workers are able to collectively reject tasks if they feel the price is unfairly low for the requested level. When a worker rejects the task from their own feed, the task is no longer available for that worker, and the requester is sent a notice. However, when more than a small percentage of workers from a level decide to reject a task, then the task is removed from everybody's task feed in the guild. With this design, requesters are incentivized to price their tasks favorably enough to match the expectations of the worker community and not underprice their tasks. In our current implementation, 3\% of workers rejecting a task removes it from the platform. 3\% was chosen arbitrarily and we hope to consider ways to optimize this value in the future.

\subsubsection{Open forum}
Forums are widely used by the crowdsourcing community. However, many of the most popular ones are external to the platforms they serve, such as Turker Nation and the /r/mturk subreddits~\cite{martin2014being}. The crowd guild forum is linked from within Daemo platform: ``Discuss'' links are available within each task to enable workers to quickly begin discussions focused on specific issues. Workers' reputation levels are publicized with badges, and separate badges are provided identifying requesters and administrators, affording direct communication and helping to humanize the platform.

\begin{figure}[tb]
 \centering
 \includegraphics[width=\columnwidth]{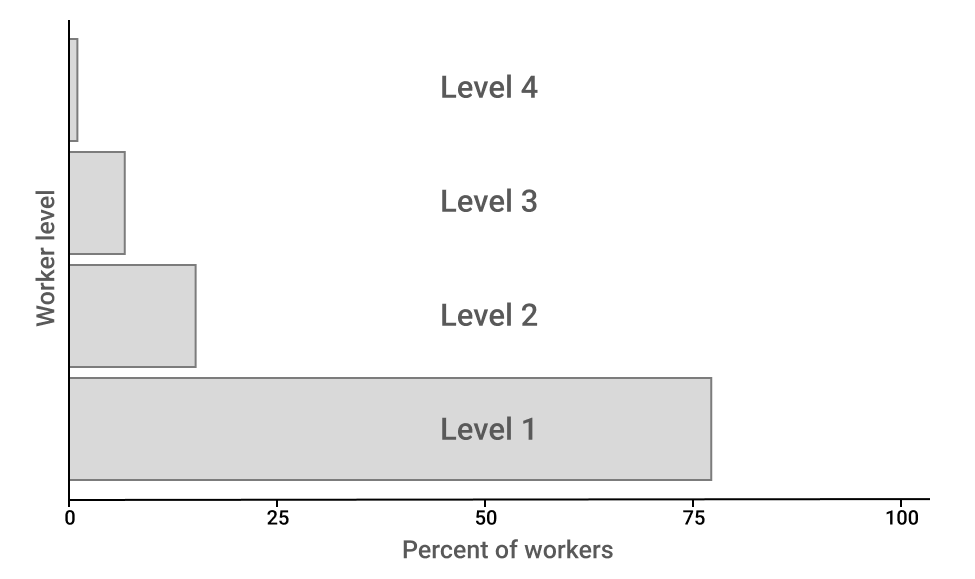}
 \caption{By the conclusion of the study, the crowd guild had evaluated its workers and distributed them across four levels.}
 \label{fig:GuildLevelDistribution}
\end{figure}

\section{Results}
After two weeks in the study, workers in the crowd guild condition spread themselves into Levels 1 to 4, with 151 of workers at Level 1, 30 in Level 2, 13 in Level 3 and 2 in Level 4 (Figure~\ref{fig:GuildLevelDistribution}). During this time, workers in the guild condition completed a total of $15176$ tasks, with an average of 87 tasks per worker. Workers in the control condition completed a total of $13427$ tasks, with an average of 113 tasks per worker.

\subsection{Crowd guilds generate accurate reputation signals}
We analyzed the relationship between workers' ground-truth accuracy and the reputation signals accessible to the system. Table~\ref{table:ModelResultsTable} contains the estimate, std. error, and p-values for the associated variables (\textit{adj.} $R^2 = 0.038$). Mean peer assessment rating was a significant predictor of ground truth accuracy ($\beta=4.1$, $p<0.005$), supporting the hypothesis that crowds can collectively author accurate reputational signals. Traditional Mechanical Turk reputation signals were not significantly correlated with ground truth accuracy, nor were the self-reported guild feedback usefulness, nor the forum activity level (all~$p>0.05$). Study condition (coded as 1 for guilds) trended toward significance $p=0.07$, suggesting that the workers in the guilds condition may have produced higher-quality results on average.
\begin{table}[tb]
\centering
\begin{tabular*}{\columnwidth}{@{\extracolsep{\fill} } l r r r r}
Coefficient & Value & Error & t-value & p-value\\
\hline
Approval rate & 1.98 & 240 & -0.59 & 0.55 \\
Accepted tasks & 0.00 & 0.00 & 0.21 & 0.83 \\
Mean peer assessment & 4.08 & 1.39 & 2.93 & \textbf{0.004} \\
Feedback usefulness & -0.34 & 0.75 & -0.45 & 0.65 \\
Forum activity & 0.09 & 0.09 & 1.01 & 0.31 \\
Condition & 3.00 & 1.66 & 1.81 & 0.07 \\
\end{tabular*}
\caption {A regression predicting workers' ground truth accuracy uncovered significant effects of average peer review score (1--4), verifying that continuous peer review can be used to establish accurate reputation credentials for workers.}
\label{table:ModelResultsTable}
\end{table}

\begin{figure}[tb]
 \centering
\includegraphics[width=.9\columnwidth]{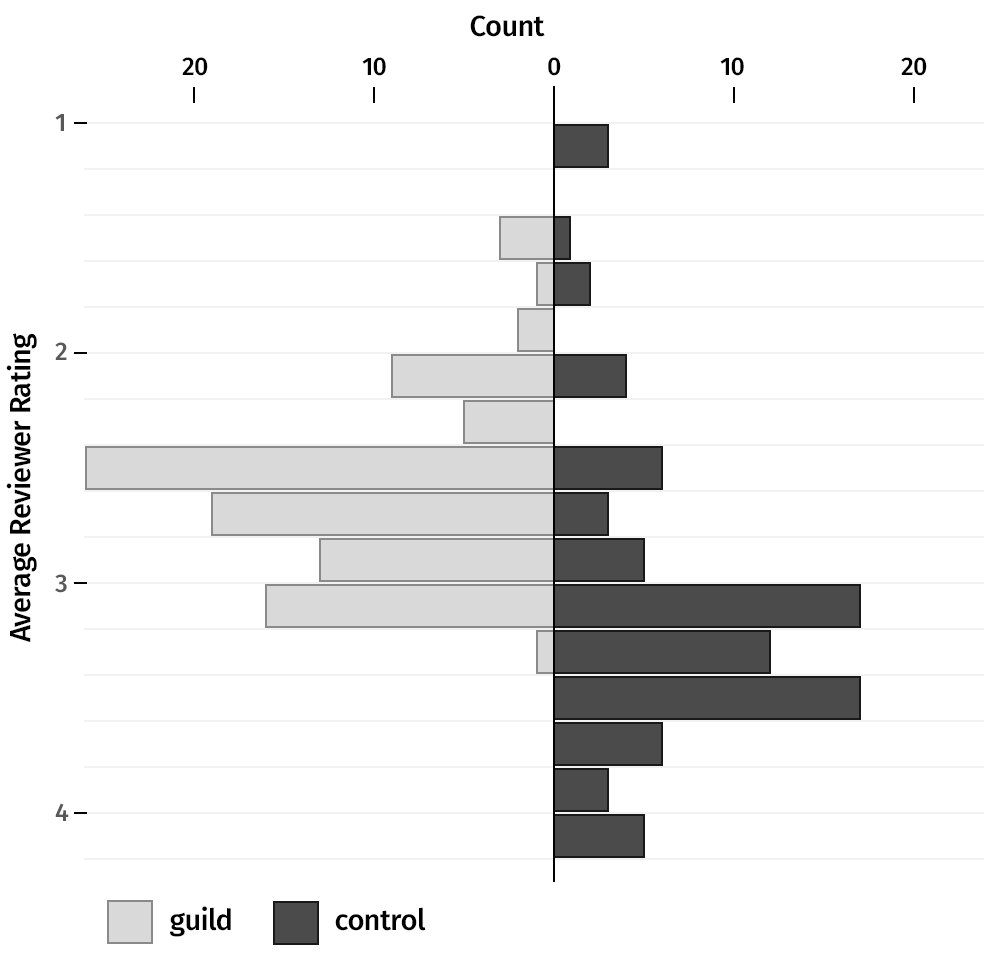}
 \caption{Workers' peer assessment ratings in the guild condition were less inflated than those in the control condition at the end of the study.}
 \label{fig:AvgReviewRatingTreatment}
\end{figure}

Was there a difference in the accuracy of peer feedback between conditions? The distribution of the overall peer assessment ratings of each worker at the conclusion of the study is shown in Figure~\ref{fig:AvgReviewRatingTreatment}. Testing correlations between workers' ground truth accuracies and their average peer assessment ratings, there was a significant correlation in the guilds condition ($p<0.001$, $r=.36$) and no significant correlation in the control condition ($p=0.09$, $r=.18$). This result suggests that feedback in the guild condition was more informative, again supporting the hypothesis that the guilds condition produced more accurate reputational information than the control condition.

To understand why crowd guilds produced more accurate reputation scores, we address two questions. First, what differed about the reputation scores between conditions? There is a statistically significant difference in the overall peer assessment ratings between conditions ($t(130.12)= 6.33$, $p<0.001$), with reviews in the guild condition ($\mu=2.5$) significantly less inflated than those in the control condition ($\mu=3.0$). Reputation inflation is a challenge in crowdsourcing marketplaces~\cite{horton2015reputation, gaikwad2016boomerang}, so a lower average score is preferable, which indicates that reviewers in the guilds condition were more discerning and spare with high ratings. Why does this difference occur? We hypothesize that workers in each condition interpreted the meaning of each rating scale level differently based on the perceived impact of ratings. This rating deflation and discernment was the likely mechanism behind the existence of a correlation between peer assessment and accuracy in the guild condition but not the control condition: in the control condition, score inflation caused workers of different ground-truth accuracies to have similar average feedback scores.

The second question: what about the crowd guilds design was responsible for producing more accurate ratings than the control condition? Potential hypotheses include the content of the free-text feedback, engagement on the forums, or the ratings themselves. The self-reported feedback usefulness and forum activity variables in the regression provide some insight into this mechanism. Neither feedback usefulness nor forum activity were significant predictors of worker accuracy (both~$p>.05$). This result suggests that the community significance of crowd guild ratings, the only other major feature in this system, was the main influencer leading to improved rating accuracy. In other words, current evidence suggests that crowd guilds produce more accurate reputation signals because workers treat the ratings differently when the ratings determine other members' levels. The guild's social structures and textual feedback may produce other benefits (e.g., a collective identity, ideas for improvement), but they do not appear to directly influence rating accuracy.

Cumulatively, these results support the hypothesis that crowd guilds produce more accurate reputation information than no guilds, and furthermore that they produce more accurate reputation information than current signals on Mechanical Turk.

\subsection{Assessing crowd guilds' qualitative impacts}
The results so far have investigated whether crowd guilds produce more accurate reputational information. However, our study also included other prototype community and feedback mechanisms, and it is important to understand the guilds' effect on the \textit{community} as well as on the reputation system. In this section, we report a mixed-methods analysis of the feedback, forums, and survey results to understand the differences in workers' community behavior between conditions. We observed that incentives pragmatize workers' advice-giving behaviour, that community integration can have strategic values for workers and requesters, and that at least some workers argue that decentralization is a benefit of crowdsourcing and should not be meddled with. Here we focus on qualitative observations made during the study on each condition's community forum and peer assessment feedback. 

\subsubsection{Crowd guilds pragmatize feedback}
Workers in both the treatment and control communities were mutually supportive. While conversations involving moderators tended to relate to bug reports or bringing up specific issues in a task, nearly all other threads were focused specifically on sharing know-how or awareness of platform features. In the guilds condition, workers often focused on more pragmatic feedback on strategies for instrumental and informational support, while control workers were more likely to offer emotional support rather than information. For example, when one worker in the control condition raised issues they were facing, another control worker empathized and replied: 

\begin{displayquote}\textit{I'm sorry. hugs I hope your day gets better. Maybe there is something the people running the study can do to fix the problem.} \end{displayquote}

On the other hand, guild workers responded to similar situations with pragmatic advice that focused on the challenge being faced. When a worker expressed concern that review tasks were not available to everyone, the response provided direct informational support, for example: 

\begin{displayquote}\textit{I have been able to do review tasks since being upped to level 2 /shrug} \end{displayquote}

This difference was also seen in the peer review feedback as part of task reviews. Reviews in the guild condition were fewer characters on average ($\mu=58.71$, $\sigma=76.94$ vs. $\mu=68.66$, $\sigma=69.60$ in control, Kruskal-Wallis $p<0.001$) and generally more focused and pragmatic. In the guild condition, responses used to let a worker know that they had done part of a task incorrectly were very direct: 

\begin{displayquote}\textit{Did not appropriately answer the question by specifying how often they used each tool.} \end{displayquote}

In the control condition, review responses exhibited doubt and consideration about the worker's thinking:

\begin{displayquote}\textit{The page doesn't say or give any clue other than people growing marijuana as to whether they are pro-legalization or not. Because of this, I personally would have marked neither. But, then again, I could be wrong. Marking neither is what I would have done to improve it.} \end{displayquote}

This difference, when considered in connection to the improved accuracy of reviews in the treatment condition mentioned above, suggests that crowd guilds improve the effectiveness of peer review by professionalizing it. In contrast, we had expected that guilds would result in a more supportive environment in their peer review. When the community was filled with workers who have control over each other's reputation, the community behaviors became more pragmatic. 

\subsubsection{Love it or hate it: workers' preferences about centralization}
Crowd guilds offer improved worker credentialing and feedback through re-centralization of reputation. In general, workers on the forum appreciated and favored these goals. A few workers in the guild condition, however, indicated a preference for an entirely decentralized platform. These ``lone wolf'' workers valued the independence of crowd work and being masters of their own fate. Lone wolf workers would often criticize guild features such as review and leveling, on the basis that they would remove independence from the worker. In our formative design feedback from crowd workers on Mechanical Turk forums, this sentiment also arose amongst a few members of the community. However, the significant majority of workers in the community expressed high levels of interest in this platform becoming a reality. We thus hypothesize that workers who have found success on Mechanical Turk as it is today see increased interdependence as a threat to their livelihood, stability and freedom.

The most salient complaints from workers in the guild condition were that they did not necessarily trust the reviewers and the leveling systems to serve the interests of the community: 

\begin{displayquote}\textit{It seems to me that people are overly judgmental in order to show that they deserve to be in Level 2 rather than in worrying about whether the rest of us should be.} \end{displayquote}

However, our quantitative analysis suggests that reviews on the treatment condition outperform the control condition in predicting actual quality of work. 

Crowd guilds replace an \textit{algorithm} (acceptance rate) with a \textit{social process} (peer assessment), and that social processes can legitimately trigger disagreements and concerns between community members. It replaces an ``us vs. them'' dynamic (workers vs. requesters) with an ``us vs. us'' dynamic (Level~1 workers vs. Level~2 reviewers). These interdependencies between workers can be good---they enable high-quality reputation signals and community feedback, for instance---but they can also cause strife and disagreements. However, if the guild develops clear metrics, norms and accountability processes over time, it can overcome many of these issues, and based on our qualitative analysis, the benefit far outweighs the cost.

\section{Discussion}
In this section, we reflect on the methodological limitations of our experiment, the major design challenges in organizing a crowd guild for a paid crowdsourcing platform, and the future directions of guilds in the crowdsourcing design space.

\subsection{Limitations}
The experiment suggests that crowd guilds enable workers to collectively manage their reputation metrics. Our evaluation was able to populate a guild with 104 workers, but large-scale crowdsourcing platforms have tens of thousands of active workers. We are unable to observe how crowd guild dynamics would play out at this scale, or over very long time periods, although crowd workers have been shown to maintain consistent performance over time~\cite{hata2016glimpse}. These efforts remain future work. It is possible that the most effective route will be to allow workers to create their own guilds, and join as many as they wish, to avoid massive guilds that are not differentiable.

External validity is also limited by Daemo's existence as a small-scale crowdsourcing platform today. We paid workers to join the evaluation; we cannot know for certain that workers would actually prefer a crowd guild like on Daemo to their existing platforms. For crowd guilds to succeed, high-quality workers and requesters must remain in the social system for long periods of time and thus have a stake in making the social system useful. An ideal evaluation would involve workers and requesters who hold such a stake, either by changing Amazon Mechanical Turk, or recruiting long-term workers and requesters onto the Daemo platform. Since Daemo is not fully evolved to conduct the entire functional experiment with real requesters and does not already have a workforce, we recruited users from Mechanical Turk. This decision made study less realistic; however, because it minimized participants' long term motivations, it is also biased toward a conservative estimate of the strength of the effect. There are possible novelty effects which will need to be teased out in future work via longitudinal studies and qualitative analysis. Finally, although we tried to recruit a wide variety of real crowd workers with real tasks from crowd marketplaces, our results might not yet generalize to crowd work at large.

\subsection{The ramifications of crowd guilds}
Introducing a sociotechnical system such as crowd guilds will inevitably shift the social and power dynamics of the crowdsourcing ecosystem. Some of these shifts are already visible. For example, the study made clear that guilds shifted the forum away from just being a water cooler and toward a work environment. In addition, the lone wolf workers were less enthused about the prospect of their reputation riding on other workers' behaviors.

We might extrapolate from these visible shifts to ones that may occur in the longer term. There will clearly be instances where workers' peer evaluations are inaccurate, and these unfair ratings may sow distrust within the guild. Over time, the guild levels may ossify and become oligarchical~\cite{shaw2014laboratories}, further making workers feel distrustful of each other. It remains to be seen whether the meta-reviews, or the ability to split off and form a separate guild, might keep these pressures in balance.

Finally, while we focused our prototype on guilds as a vehicle for reputation, they could grow to encompass other functions as well. We prototyped several of these functions---collective rejection and wage setting, for example. We intend these prototypes to stand as example guild functions based on the roles that previous guilds played. However, crowd guilds may introduce entirely new roles not seen in historical guilds. Could they produce new forms of training? New social environments? Stronger relationships amongst crowd workers?
 
\subsection{Future work}
The future of crowd work depends on strong worker motivation, feedback, and pay~\cite{kittur2013future}. There exist many other mechanisms for achieving these goals, especially increasing a sense of belonging instead of isolation. Crowd guilds begin with a narrow goal and pragmatic design, aimed to provide mechanisms for stronger reputation for workers and thus fairer pay on the platform. A second step is to translate this centralization into increased worker collective action in solving problems such as asymmetric access to information, limitations on open innovation and governance problems within the online labour marketplace~\cite{salehi2015we}. More broadly, we believe that effective social computing designs can go further in enabling more prosocial and equitable environments for crowd workers. One worker mentioned:

\begin{displayquote}\textit{I was just thinking that it would be very helpful/useful to have some type of notification that signifies if a Forum Member is Currently Active/On or if they are Away etc.. If we knew that specific Members were Online, we would feel more connected and have a sense that we have someone to turn to in real time.}\end{displayquote}

Just as forum designs affect the trust in crowd worker forums~\cite{laplante2016building}, it is important to consider how guilds design will influence governance. How will crowd guilds govern themselves long term? The guild in our experiment had no leadership structure, but if it were to persist, it would need to develop one. What policies can they control, how do they make decisions, and how do they collect and redistribute their own income? These questions combine social computing design and political science. In the future, we hope to analyze variation in guild governance policies to better understand the forms of self-governance that predict long-term engagement and satisfaction.

Finally, in the current design, the guild represents all the workers on the platform: where every worker is a part of the guild, and they all collectively assess everyone's work. However, reviewers may not have the domain knowledge in every subject area to produce a quality review. We envision that different guilds will form around particular communities of practice.

\section{Conclusion}
Crowd workers in microtask platforms have been decentralized in order to reap efficiencies from independent work. However, in the long term, it is crucial for workers to have opportunities to connect with each other, learn from each other, and impact the platforms they use. In order to address this, we have drawn upon the historical example of guilds to bring workers into a loose affiliation that can certify each other's quality. Thus, this paper introduces crowd guilds, a system that empowers worker communities with peer assessed reviews with feedback leveling, and a connected community forum. Our evaluation of crowd guilds demonstrated that crowd guilds lead to improved reputation signals and community behaviour shifts toward efficient feedback. More generally, crowd guilds offer opportunities to co-design crowdsourcing platforms with worker platforms.

\section{Acknowledgement}
We thank the following members of the Stanford Crowd Research Collective for their contributions: Alison Cossette, Trygve Cossette, Ryan Compton, Pierre Fiorini, Flavio S Nazareno, Karolina Ziukloski, Lucas Bamidele, Durim Morina, Sharon Zhou, Senadhipathige Suminda Niranga, Gagana B, Jasmine Lin, William Dai, Leonardy Kristianto, Samarth Sandeep, Justin Le, Manoj Pandey, Vinayak Mathur, Kamila Mananova, Ahmed Nasser, Preethi Srinivas, Sachin Sharma, Christopher Diemert, Lakmal Rupasinghe, Aditi Mithal, Divya Nekkanti, Mahesh Murag, Alan James, Vrinda Bhatia, Venkata Karthik Gullapalli, Yen Huang, Armine Rezaiean-Asel, Aditya V. Nadimpalli, Jaspreet Singh Riar, Rohheit Nistala, Mathias Burton, Anuradha Jayakody, Vinet Sethia, Yash Mittal, Victoria Purynova, Shrey Gupta, Alex Stolzoff, Paul Abhratanu, Carl Chen, Namit Juneja, Jorg Nathaniel Doku, Anmol Agarwal, Chen Xi, Zain Rehmani, Yashovardhan Sharma, Yash Sherry, Sree Nihit Munakala, Shivi Bajpai, Shivam Agarwal, Sanjiv Lobo, Raymond Su, Prithvi Raj, Nikita Dubnov, Kevin Viet Le, Klerisson Paixao, Haritha Thilakarathne, David Thompson, Varun Hasija, Lucas Qiu, Kushagro Bhattacharjee, Karthik Paga, Ishan Yelurwar, Archana Dhankar, Aalok Thakkar. We also thank the Amazon Mechanical Turk workers who participated in this study. This work was supported by NSF award IIS-1351131, Office of Naval Research award N00014-16-1-2894, Toyota and the Hasso Plattner Design Thinking Research Program.

\balance{}

\bibliographystyle{acm-sigchi-msb}
\bibliography{ogov_cscw_paper}

\begin{thebibliography}{10}
\providecommand{\url}[1]{\texttt{#1}}
\providecommand{\urlprefix}{URL }

\bibitem{anderson2011crowdsourcing}
Anderson, M.
\newblock Crowdsourcing higher education: A design proposal for distributed
  learning.
\newblock \emph{MERLOT Journal of Online Learning and Teaching}, 7(4):576--590,
  2011.

\bibitem{billett2001learning}
Billett, S.
\newblock \emph{Learning in the workplace: Strategies for effective practice.}
\newblock ERIC, 2001.

\bibitem{boud2000sustainable}
Boud, D.
\newblock Sustainable assessment: rethinking assessment for the learning
  society.
\newblock \emph{Studies in continuing education}, 22(2):151--167, 2000.

\bibitem{boud2013enhancing}
Boud, D. et~al.
\newblock \emph{Enhancing learning through self-assessment}.
\newblock Routledge, 2013.

\bibitem{callison2009fast}
Callison-Burch, C.
\newblock Fast, cheap, and creative: evaluating translation quality using
  amazon's mechanical turk.
\newblock In \emph{Proceedings of the 2009 Conference on Empirical Methods in
  Natural Language Processing: Volume 1-Volume 1}, pp. 286--295. Association
  for Computational Linguistics, 2009.

\bibitem{campbell2015thousands}
Campbell, J.A., et~al.
\newblock Thousands of positive reviews: Distributed mentoring in online fan
  communities.
\newblock \emph{arXiv preprint arXiv:1510.01425}, 2015.

\bibitem{cheng2015measuring}
Cheng, J., Teevan, J., and Bernstein, M.S.
\newblock Measuring crowdsourcing effort with error-time curves.
\newblock In \emph{Proceedings of the 33rd Annual ACM Conference on Human
  Factors in Computing Systems}, pp. 1365--1374. ACM, 2015.

\bibitem{deng2013crowdsourcing}
Deng, X.N. and Joshi, K.
\newblock Is crowdsourcing a source of worker empowerment or exploitation?
  understanding crowd workers' perceptions of crowdsourcing career.
\newblock 2013.

\bibitem{dontcheva2014combining}
Dontcheva, M., Morris, R.R., Brandt, J.R., and Gerber, E.M.
\newblock Combining crowdsourcing and learning to improve engagement and
  performance.
\newblock In \emph{Proceedings of the 32nd annual ACM conference on Human
  factors in computing systems}, pp. 3379--3388. ACM, 2014.

\bibitem{dow2012shepherding}
Dow, S., Kulkarni, A., Klemmer, S., and Hartmann, B.
\newblock Shepherding the crowd yields better work.
\newblock In \emph{Proceedings of the ACM 2012 conference on Computer Supported
  Cooperative Work}, pp. 1013--1022. ACM, 2012.

\bibitem{ducheneaut2007life}
Ducheneaut, N., Yee, N., Nickell, E., and Moore, R.J.
\newblock The life and death of online gaming communities: a look at guilds in
  world of warcraft.
\newblock In \emph{Proceedings of the SIGCHI conference on Human factors in
  computing systems}, pp. 839--848. ACM, 2007.

\bibitem{falchikov2000student}
Falchikov, N. and Goldfinch, J.
\newblock Student peer assessment in higher education: A meta-analysis
  comparing peer and teacher marks.
\newblock \emph{Review of educational research}, 70(3):287--322, 2000.

\bibitem{gaikwad2016boomerang}
Gaikwad, S., et~al.
\newblock Boomerang: Rebounding the consequences of reputation feedback on
  crowdsourcing platforms.
\newblock In \emph{Proceedings of the 29th Annual Symposium on User Interface
  Software and Technology}, pp. 625--637. ACM, 2016.

\bibitem{gaikwad2015daemo}
Gaikwad, S.N., et~al.
\newblock Daemo: A self-governed crowdsourcing marketplace.
\newblock In \emph{Proceedings of the 28th Annual ACM Symposium on User
  Interface Software \& Technology}, pp. 101--102. ACM, 2015.

\bibitem{gilbert2014if}
Gilbert, E.
\newblock What if we ask a different question?: social inferences create
  product ratings faster.
\newblock In \emph{Proceedings of the SIGCHI Conference on Human Factors in
  Computing Systems}, pp. 2759--2762. ACM, 2014.

\bibitem{gray2016crowd}
Gray, M.L., Suri, S., Ali, S.S., and Kulkarni, D.
\newblock The crowd is a collaborative network.
\newblock \emph{Proceedings of Computer-Supported Cooperative Work}, 2016.

\bibitem{gupta2014turk}
Gupta, N., Martin, D., Hanrahan, B.V., and O'Neill, J.
\newblock Turk-life in india.
\newblock In \emph{Proceedings of the 18th International Conference on
  Supporting Group Work}, pp. 1--11. ACM, 2014.

\bibitem{guthrie2007learning}
Guthrie, C.
\newblock On learning the research craft: Memoirs of a journeyman researcher.
\newblock \emph{Journal of Research Practice}, 3(1):1, 2007.

\bibitem{haas2015argonaut}
Haas, D., Ansel, J., Gu, L., and Marcus, A.
\newblock Argonaut: macrotask crowdsourcing for complex data processing.
\newblock \emph{Proceedings of the VLDB Endowment}, 8(12):1642--1653, 2015.

\bibitem{hata2016glimpse}
Hata, K., Krishna, R., Fei-Fei, L., and Bernstein, M.S.
\newblock A glimpse far into the future: Understanding long-term crowd worker
  accuracy.
\newblock \emph{arXiv preprint arXiv:1609.04855}, 2016.

\bibitem{horton2015reputation}
Horton, J. and Golden, J.
\newblock Reputation inflation: Evidence from an online labor market.
\newblock \emph{Work. Pap., NYU}, 2015.

\bibitem{irani2013turkopticon}
Irani, L.C. and Silberman, M.
\newblock Turkopticon: Interrupting worker invisibility in amazon mechanical
  turk.
\newblock In \emph{Proceedings of the SIGCHI Conference on Human Factors in
  Computing Systems}, pp. 611--620. ACM, 2013.

\bibitem{josang2007survey}
J{\o}sang, A., Ismail, R., and Boyd, C.
\newblock A survey of trust and reputation systems for online service
  provision.
\newblock \emph{Decision support systems}, 43(2):618--644, 2007.

\bibitem{kang2009impact}
Kang, J., Ko, I., and Ko, Y.
\newblock The impact of social support of guild members and psychological
  factors on flow and game loyalty in mmorpg.
\newblock In \emph{System Sciences, 2009. HICSS'09. 42nd Hawaii International
  Conference on}, pp. 1--9. IEEE, 2009.

\bibitem{karoly200421st}
Karoly, L.A. and Panis, C.W.
\newblock \emph{The 21st century at work: Forces shaping the future workforce
  and workplace in the United States}, vol. 164.
\newblock Rand Corporation, 2004.

\bibitem{kaye2012putting}
Kaye, L.K. and Bryce, J.
\newblock Putting the fun factor into gaming: The influence of social contexts
  on the experiences of playing videogames.
\newblock \emph{International Journal of Internet Science}, 7(1):24--38, 2012.

\bibitem{kieser1989organizational}
Kieser, A.
\newblock Organizational, institutional, and societal evolution: Medieval craft
  guilds and the genesis of formal organizations.
\newblock \emph{Administrative Science Quarterly}, pp. 540--564, 1989.

\bibitem{kittur2013future}
Kittur, A., et~al.
\newblock The future of crowd work.
\newblock In \emph{Proceedings of the 2013 conference on Computer supported
  cooperative work}, pp. 1301--1318. ACM, 2013.

\bibitem{kulkarni2012mobileworks}
Kulkarni, A., et~al.
\newblock Mobileworks: Designing for quality in a managed crowdsourcing
  architecture.
\newblock \emph{Internet Computing, IEEE}, 16(5):28--35, 2012.

\bibitem{kulkarni2015peer}
Kulkarni, C., et~al.
\newblock Peer and self assessment in massive online classes.
\newblock \emph{ACM Transactions on Computer-Human Interaction (TOCHI)},
  20(6):33, 2013.

\bibitem{kulkarni2015peerstudio}
Kulkarni, C.E., Bernstein, M.S., and Klemmer, S.R.
\newblock Peerstudio: Rapid peer feedback emphasizes revision and improves
  performance.
\newblock In \emph{Proceedings of the Second (2015) ACM Conference on Learning@
  Scale}, pp. 75--84. ACM, 2015.

\bibitem{laplante2016building}
LaPlante, R. and Silberman, M.S.
\newblock Building trust in crowd worker forums: Worker ownership, governance,
  and work outcomes.
\newblock In \emph{Proceedings of WebSci16}. ACM, 2016.

\bibitem{larson1979rise}
Larson, M.S. and Larson, M.S.
\newblock \emph{The rise of professionalism: A sociological analysis}, vol.
  233.
\newblock Univ of California Press, 1979.

\bibitem{laubacher1997flexible}
Laubacher, R.J., Malone, T.W., et~al.
\newblock Flexible work arrangements and 21st century worker's guilds.
\newblock Tech. rep., MIT Center for Coordination Science, 1997.

\bibitem{lee2003knowledge}
Lee, H. and Choi, B.
\newblock Knowledge management enablers, processes, and organizational
  performance: An integrative view and empirical examination.
\newblock \emph{Journal of management information systems}, 20(1):179--228,
  2003.

\bibitem{leskovec2010signed}
Leskovec, J., Huttenlocher, D., and Kleinberg, J.
\newblock Signed networks in social media.
\newblock In \emph{Proceedings of the SIGCHI conference on human factors in
  computing systems}, pp. 1361--1370. ACM, 2010.

\bibitem{little2010exploring}
Little, G., Chilton, L.B., Goldman, M., and Miller, R.C.
\newblock Exploring iterative and parallel human computation processes.
\newblock In \emph{Proceedings of the ACM SIGKDD workshop on human
  computation}, pp. 68--76. ACM, 2010.

\bibitem{little2010turkit}
Little, G., Chilton, L.B., Goldman, M., and Miller, R.C.
\newblock Turkit: human computation algorithms on mechanical turk.
\newblock In \emph{Proceedings of the 23nd annual ACM symposium on User
  interface software and technology}, pp. 57--66. ACM, 2010.

\bibitem{luther2015structuring}
Luther, K., et~al.
\newblock Structuring, aggregating, and evaluating crowdsourced design
  critique.
\newblock In \emph{Proceedings of the 18th ACM Conference on Computer Supported
  Cooperative Work \& Social Computing}, pp. 473--485. ACM, 2015.

\bibitem{malone1999will}
Malone, T.W. and Laubacher, R.J.
\newblock How will work change? elancers, empowerment, and guilds.
\newblock \emph{THE PROMISE OF GLOBAL NETWORKS}, p. 119, 1999.

\bibitem{martin2014being}
Martin, D., Hanrahan, B.V., O'Neill, J., and Gupta, N.
\newblock Being a turker.
\newblock In \emph{Proceedings of the 17th ACM conference on Computer supported
  cooperative work \& social computing}, pp. 224--235. ACM, 2014.

\bibitem{mcinnis2016taking}
McInnis, B., Cosley, D., Nam, C., and Leshed, G.
\newblock Taking a hit: Designing around rejection, mistrust, risk, and
  workers' experiences in amazon mechanical turk.
\newblock In \emph{Proceedings of the 2016 CHI Conference on Human Factors in
  Computing Systems}, pp. 2271--2282. ACM, 2016.

\bibitem{mitra2015comparing}
Mitra, T., Hutto, C.J., and Gilbert, E.
\newblock Comparing person-and process-centric strategies for obtaining quality
  data on amazon mechanical turk.
\newblock In \emph{Proceedings of the 33rd Annual ACM Conference on Human
  Factors in Computing Systems}, pp. 1345--1354. ACM, 2015.

\bibitem{mocarelli2008guilds}
Mocarelli, L.
\newblock Guilds reappraised: Italy in the early modern period.
\newblock \emph{International review of social history}, 53(S16):159--178,
  2008.

\bibitem{ogilvie2004guilds}
Ogilvie, S.
\newblock Guilds, efficiency, and social capital: evidence from german
  proto-industry.
\newblock \emph{Economic history review}, pp. 286--333, 2004.

\bibitem{ogilvie2005use}
Ogilvie, S.
\newblock The use and abuse of trust: the deployment of social capital by early
  modern guilds.
\newblock \emph{Jahrbuch f{\"u}r Wirtschaftsgeschichte}, 1:15--52, 2005.

\bibitem{patel2012power}
Patel, N., et~al.
\newblock Power to the peers: authority of source effects for a voice-based
  agricultural information service in rural india.
\newblock In \emph{Proceedings of the Fifth International Conference on
  Information and Communication Technologies and Development}, pp. 169--178.
  ACM, 2012.

\bibitem{peng2010decision}
Peng~Dai, M.D. and Weld, S.
\newblock Decision-theoretic control of crowd-sourced workflows.
\newblock In \emph{In the 24th AAAI Conference on Artificial Intelligence
  (AAAI'10)}. Citeseer, 2010.

\bibitem{perez2008inventing}
P{\'e}rez, L.
\newblock Inventing in a world of guilds: silk fabrics in eighteenth-century
  lyon.
\newblock \emph{Guilds, innovation, and the European economy}, pp. 1400--1800,
  2008.

\bibitem{poor2015mmo}
Poor, N.
\newblock What mmo communities don't do: A longitudinal study of guilds and
  character leveling, or not.
\newblock In \emph{Ninth International AAAI Conference on Web and Social
  Media}. 2015.

\bibitem{rees2015building}
Rees~Lewis, D., Harburg, E., Gerber, E., and Easterday, M.
\newblock Building support tools to connect novice designers with professional
  coaches.
\newblock In \emph{Proceedings of the 2015 ACM SIGCHI Conference on Creativity
  and Cognition}, pp. 43--52. ACM, 2015.

\bibitem{renard1918guilds}
Renard, G.
\newblock Guilds in the middle ages.
\newblock 1918.

\bibitem{rosser1997crafts}
Rosser, G.
\newblock Crafts, guilds and the negotiation of work in the medieval town.
\newblock \emph{Past \& Present}, (154):3--31, 1997.

\bibitem{rzeszotarski2011instrumenting}
Rzeszotarski, J.M. and Kittur, A.
\newblock Instrumenting the crowd: using implicit behavioral measures to
  predict task performance.
\newblock In \emph{Proceedings of the 24th annual ACM symposium on User
  interface software and technology}, pp. 13--22. ACM, 2011.

\bibitem{salehi2015we}
Salehi, N., et~al.
\newblock We are dynamo: Overcoming stalling and friction in collective action
  for crowd workers.
\newblock In \emph{Proceedings of the 33rd Annual ACM Conference on Human
  Factors in Computing Systems}, pp. 1621--1630. ACM, 2015.

\bibitem{shaw2014laboratories}
Shaw, A. and Hill, B.M.
\newblock Laboratories of oligarchy? how the iron law extends to peer
  production.
\newblock \emph{Journal of Communication}, 64(2):215--238, 2014.

\bibitem{sheng2008get}
Sheng, V.S., Provost, F., and Ipeirotis, P.G.
\newblock Get another label? improving data quality and data mining using
  multiple, noisy labelers.
\newblock In \emph{Proceedings of the 14th ACM SIGKDD international conference
  on Knowledge discovery and data mining}, pp. 614--622. ACM, 2008.

\bibitem{sparrowe2001social}
Sparrowe, R.T., Liden, R.C., Wayne, S.J., and Kraimer, M.L.
\newblock Social networks and the performance of individuals and groups.
\newblock \emph{Academy of management journal}, 44(2):316--325, 2001.

\bibitem{surowiecki2005wisdom}
Surowiecki, J.
\newblock \emph{The wisdom of crowds}.
\newblock Anchor, 2005.

\bibitem{suzuki2016atelier}
Suzuki, R., et~al.
\newblock Atelier: Repurposing expert crowdsourcing tasks as micro-internships.
\newblock \emph{arXiv preprint arXiv:1602.06634}, 2016.

\bibitem{topping1998peer}
Topping, K.
\newblock Peer assessment between students in colleges and universities.
\newblock \emph{Review of educational Research}, 68(3):249--276, 1998.

\bibitem{ulibarri2014research}
Ulibarri, N., et~al.
\newblock Research as design: Developing creative confidence in doctoral
  students through design thinking.
\newblock \emph{International Journal of Doctoral Studies}, 9:249--270, 2014.

\bibitem{wenger2011communities}
Wenger, E.
\newblock Communities of practice: A brief introduction.
\newblock 2011.

\bibitem{yin2016communication}
Yin, M., Gray, M.L., Suri, S., and Vaughan, J.W.
\newblock The communication network within the crowd.
\newblock In \emph{Proceedings of the 25th International Conference on World
  Wide Web}, pp. 1293--1303. International World Wide Web Conferences Steering
  Committee, 2016.

\bibitem{yu2014comparison}
Yu, L., Andr{\'e}, P., Kittur, A., and Kraut, R.
\newblock A comparison of social, learning, and financial strategies on crowd
  engagement and output quality.
\newblock In \emph{Proceedings of the 17th ACM conference on Computer supported
  cooperative work \& social computing}, pp. 967--978. ACM, 2014.

\bibitem{zhu2014reviewing}
Zhu, H., Dow, S.P., Kraut, R.E., and Kittur, A.
\newblock Reviewing versus doing: Learning and performance in crowd assessment.
\newblock In \emph{Proceedings of the 17th ACM conference on Computer supported
  cooperative work \& social computing}, pp. 1445--1455. ACM, 2014.

\end{thebibliography}
\end{document}